\documentclass[11pt,a4paper]{article} 

\usepackage[T1]{fontenc}
\usepackage[utf8]{inputenc}   
\usepackage{lmodern}          
\usepackage{microtype}        
\usepackage{amsmath,amssymb,mathtools}
\usepackage{graphicx}
\usepackage{siunitx}          
\usepackage{authblk}          
\usepackage[colorlinks=true,linkcolor=blue,citecolor=blue,urlcolor=blue]{hyperref}
\usepackage[margin=2.5cm]{geometry} 
\usepackage{placeins}
\usepackage{subcaption}

\title{Quantum Late-Time Decay and Channel Dependence}

\author[1,2]{Francesco Giacosa\thanks{corresponding author: francesco.giacosa@ujk.edu.pl}}
\author[1]{Anna Kolbus}
\author[1]{Krzysztof Kyziol}
\author[1]{Magdalena Plodowska}
\author[1]{Milena Piotrowska}
\author[1]{Karol Szary}
\author[1]{Arthur Vereijken}

\affil[1]{Jan Kochanowski University, Kielce, Poland}
\affil[2]{Goethe University, Frankfurt am Main, Germany}

\begin{document}
\maketitle
\begin{abstract}
\noindent
Quantum mechanics predicts deviations from exponential decay at short and long times, yet experimental evidence is limited. We report a power-law tail after $\sim$10 lifetimes in two fluorescent compounds (erythrosine~B and eosine Y), confirmed by two detectors probing distinct bands but yielding different power coefficients. The data match a divergent but normalizable spectral density, and theory predicts oscillations as a future test. A novel and general result is that in multichannel QM (and QFT) decay, the lifetime is universal, but the late-time deviations are channel- (or band-) dependent, a feature consistent with our data.
\end{abstract}

\section{Introduction.}
The exponential law of spontaneous decay, a cornerstone of both non-relativistic quantum mechanics (QM) and relativistic quantum field theory (QFT), is expressed as $P(t)=e^{-t/\tau}$, where $P(t)$ is the survival probability and $\tau$ the lifetime of the unstable state.
However, this simple law is known to break down at very short and very long times, e.g. ~\cite{ghirardi,Facchi:2008nrb,Kofman:2000gle,dicus2002} for QM and ~\cite{Facchi:1999ik,Giacosa:2021hgl,Giacosa:2011xa} for QFT. 
At early times, $P(t)$ starts out flat (zero slope, $P'(0)=0$), while at later times it turns into a power-law tail \cite{khalfin}, $P(t)\sim t^{-(\beta-1)}$. In turn, $-P'(t)dt$ is the probability that a single unstable state decays between $t$ and $t+dt$. 
For $N_0$ initial unstable states, the measured decay rate intensity $I(t)$ (assuming that the $N_0$ states decay independently from one another, and there is no further interaction among the decay products) reads 
\begin{equation}
    I(t)=-N_0\frac{dP}{dt}
    \text{ ,}
\end{equation}
thus for large times $I(t) \propto-P'(t)\sim t^{-\beta}$.  

On the experimental side, few confirmations of these deviations exist. At short times, deviations (including both a Zeno and anti-Zeno regimes with slowed and increased decay rates) were measured in tunnelling of sodium atoms through an accelerated optical potential~\cite{Wilkinson:1997sez,ZenoUnstable}. 
At long times, deviations were seen in fluorescence decays of various chemical compounds, such as rhodamine and polyfluorene~\cite{Rothe:2006rma}. The decay rate $I(t)$ could be monitored\footnote{For fluorescence, it is assumed that each decay event produces one detected fluorescence photon and further absorption and/or rescattering are neglected: the measured fluorescence intensity is
proportional to the decay-rate intensity $I(t)$.} for a very long time with power exponents $\beta$ ranging between $2$--$4$ and its onset (the `turnover' time at which the power law starts to dominate) between $8$--$20$ lifetimes.

Besides these measurements, indirect evidence of both short- and long-time deviations was obtained from photons propagating in waveguide arrays~\cite{crespi}, where `time' is replaced by `space'. In nuclear physics, a late-time power law after $40 \tau$ with $\beta=7.4$ in the decay $^8\mathrm{Be}\rightarrow \alpha\alpha$ was predicted in Ref.~\cite{Kelkar:2004zz} by reconstructing the spectral function from $\alpha\alpha$ scattering data. 
In the realm of strong interactions, deviations are expected to be large due to large distortions from Breit-Wigner type~\cite{Giacosa:2007bn,Giacosa:2021mbz}, as confirmed e.g. by the experimental data of Ref.~\cite{ALEPH:2005qgp}. Anyway, strong decays are too fast for a direct measurement.

In general, for isolated atomic or nuclear systems, the short-time deviations occur extremely early and the long-time ones extremely late, rendering them very hard to observe. 
A good illustration is the well-known $2P\!\rightarrow\!1S$ transition of the hydrogen atom with $\tau=1.595$\,ns. 
As shown analytically~\cite{Facchi:1998abc} and numerically~\cite{Giacosa:2024yhn}, short-time deviations (with both Zeno and anti-Zeno intervals) occur on the attosecond scale\footnote{This temporal region could be tested by using short attosecond laser pulses, e.g. \cite{Bandrauk2017}.} ($10^{-9}\tau$), while late-time deviations start at $200$\,ns ($125\tau$), when basically no excited atoms remain. 
These facts explain why deviations could be observed in engineered optical potentials at short times, or in dye fluorescence with very good precision at late times. However, since the work of Rothe \emph{et al.}~\cite{Rothe:2006rma}, (to our knowledge) no independent confirmation in other systems has been reported.
 
The aim of the present work is to investigate, both experimentally and theoretically, the late-time decay. 
Molecules decaying via fluorescence provide an optimal environment for verifying late-time deviations.  To this end, we investigate the fluorescence decay in organic dyes, among which two of them 
(erythrosine~B and eosine~Y) exhibit clear deviations from the exponential behaviour at late times.
A novelty of our late-time study is the use of two detectors operating in distinct spectral windows.  While both channels report the same exponential lifetime, the coefficients of the late-time power law differ, with $\beta\approx 1.5$ and $\beta\approx 2.0$ respectively, see details below. 

This observation compelled us to examine the quantum decay law at large times in greater generality. Remarkably, the measured values of the power–law exponent $\beta$
can only arise from very specific forms of the spectral function. Even more significantly, the analysis forced us to reconsider the properties of the multichannel decay law at late times \cite{Giacosa:2021hgl,Giacosa:2011xa}, showing that quantum mechanics indeed allows for distinct power laws to emerge in different spectral bands.
In this context we also establish novel general relations, which are not restricted to a specific system but should hold for any unstable quantum state.

\section{Experiment: setup and results} 
\begin{figure}[!h]
    \centering
    \begin{subfigure}{0.48\textwidth}
        \includegraphics[width=\linewidth]{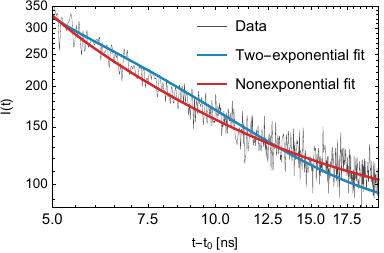}
        \label{fig:a}
        \caption{Erythrosine B, Channel 1}
    \end{subfigure}
    \hfill
    \begin{subfigure}{0.48\textwidth}
        \includegraphics[width=\linewidth]{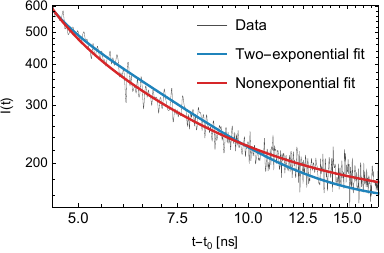}
        \label{fig:b}
        \caption{Erythrosine B, Channel 2}
    \end{subfigure}
    
    \vspace{4em}
    
    \begin{subfigure}{0.48\textwidth}
        \includegraphics[width=\linewidth]{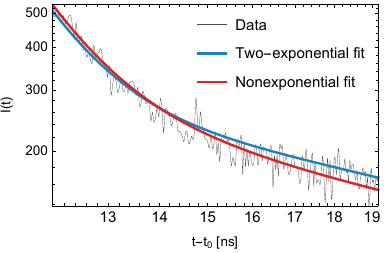}
        \label{fig:c}
        \caption{Eosine Y, Channel 1}
    \end{subfigure}
    \hfill
    \begin{subfigure}{0.48\textwidth}
        \includegraphics[width=\linewidth]{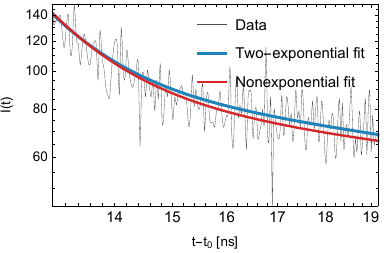}
        \label{fig:d}
        \caption{Eosine Y, Channel 2}
    \end{subfigure}
    \caption{Fluorescence intensity for both photon detectors (channel 1: left, channel 2: right) - comparison between data and both fitting functions up to $\sim 20$ ns (data taking up to $\sim 95$ ns).}
    \label{fig:non-exp_comb}
\end{figure}

The experiment was performed on a Nikon Eclipse Ti-E inverted confocal microscope equipped with two picosecond laser diodes as an excitation source (PicoQuant LDH-D-C-440 with wavelength 438 nm and PicoQuant LDH-D-C-485 with wavelength 483 nm -- pulse widths $< 120$\,ps, spectral widths ranging from 2 to 8 nm; only the former was used). 
Fluorescence from the sample was coupled via optical fibres into a dual-channel detection unit (two PicoQuant PMA Hybrid 40 hybrid detectors, timing resolution $<$120\,ps), separated by a~dichroic mirror and bandpass filters (520/35\,nm and 600/50\,nm). 
Time-correlated single-photon counting was carried out with a PicoHarp 300 module. 
This dual-detector scheme allowed us to record fluorescence decay simultaneously in two distinct spectral windows. 
The five dyes (fluorescein, acridine orange, rhodamine~B, erythrosine~B, and eosine Y) were dissolved in methanol or water with a~solution concentration of 10$^{-5}$ mol/dm$^{3}$.
While fluorescein, acridine orange, and rhodamine~B can be described by a single or two-exponential function (plus background), erythrosine~B and eosine Y exhibit a clear non-exponential behaviour at late times. Hence, we focus below on erythrosine~B and eosine Y in the following. (Results on the two-exponential decay of acridine orange can be found in \cite{Giacosa:2025AO} and in the supplemental material.)
These data sets were combined together and are reported in Fig. \ref{fig:non-exp_comb}.

We used two different fitting models presented in Table \ref{tab:my_label}: a two-exponential function and an exponential one plus a power law.  The fit results are reported in Tables \ref{data-analysis table 3} and \ref{data-analysis table 2} (for more details, see supplemental material): the non-exponential model describes the data significantly better than the two-exponential one, as the $\chi^2$ values show. This is confirmed in Fig. \ref{fig:non-exp_comb} for selected time intervals, where the exponential fit systematically underestimates (or overestimates) the data. 

Moreover, the exponential leads to inconsistent values for the second lifetime $\tau_2$ obtained in channel 1 and channel 2, respectively. This is a serious drawback for this description. On the contrary, whenever the double-exponential fit is successful, the values of $\tau_2$ are compatible (see results for acridine orange data in supplemental material). 

The coefficient $\beta$ for the power laws is also channel dependent, but this is indeed in agreement with QM, see the following. 

\begin{table}[!htbp]
    \caption[Model functions employed in the analysis.]{Model functions employed in the analysis; $t \gtrsim 2t_0$ and $b$ is the background. Note, $t_0$ is not a fit parameter, but is taken as the maximum of the intensity.}
    \centering
    {\renewcommand{\arraystretch}{2}
    \begin{tabular}{|c|c|c|}
         \hline \hline
         Model & Fluorescence intensity $I(t)$ & Fit parameters \\ \hline \hline
         Two-exponential & $I(t)=C_1\exp \left( -\frac{t-t_0}{\tau_1} \right) + C_2\exp \left( -\frac{t-t_0}{\tau_2} \right) + b$ & $\chi ^2 (C_1,\tau _1, C_2,\tau _2, b)$ \\ \hline
         Nonexponential & $I(t)=C \exp \left( -\frac{t-t_0}{\tau} \right) + C_p \, (t-t_0)^{-\beta} +b$ & $\chi ^2 (C,\tau, C_p,\beta, b)$ \\ \hline
    \end{tabular}}
    \label{tab:my_label}
\end{table}

\begin{figure}
    \centering
    \includegraphics[width=\linewidth]{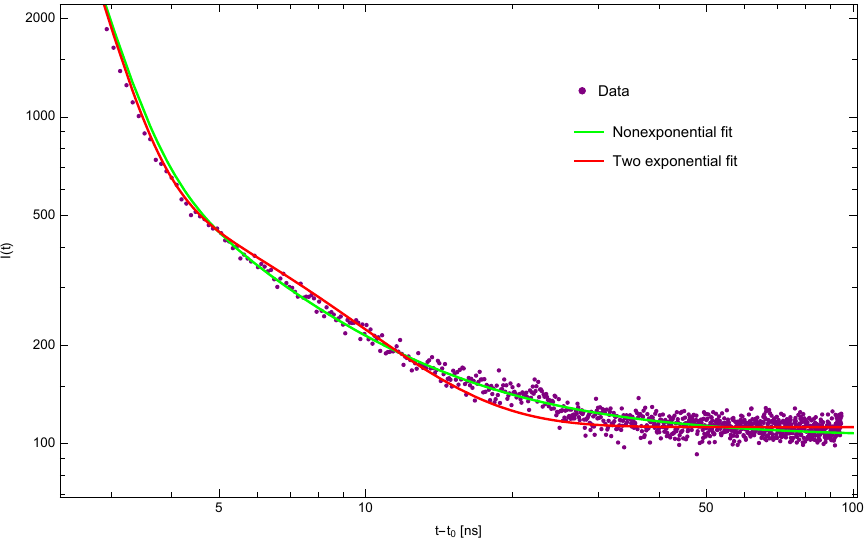}
    \caption{Comparison between two-exponential and nonexponential models on a full timescale using a 3-point moving average (Erythrosine B, Channel 1). }
    \label{fig:plots-full-timescale}
\end{figure}

For a clearer visualization of the difference between the two-exponential
and the power-law descriptions, we additionally show in Fig. \ref{fig:plots-full-timescale} a 3-point
moving average of the data for the case of channel 1, erythrosine B. Specifically, for a sequence of measured
intensities $I_k$, we define
\begin{equation}
\tilde I_k=\frac{I_{k-1}+I_k+I_{k+1}}{3}\, .
\end{equation}
This standard procedure reduces statistical fluctuations, particularly in
the low-count late-time region, while preserving the overall shape of the
decay curve. As a consequence, the superiority of the power-law description
over the two-exponential one becomes more apparent, specifically at intermediate and late times.

\begin{table}[!h]
    \centering
    \tiny
    \caption{Fitting results for Erythrosine B measurements.}
    \label{data-analysis table 3}
    \renewcommand{\arraystretch}{1.7}
    \begin{tabular}{|c|c|c|c|c|c|c|c|c|c|c|c|c|}
    \hline
    \multicolumn{13}{|c|}{Fitting Range - Channel 1: 0.960 - 94.752 ns; Channel 2: 0.960 - 94.656 ns;} \\ \hline
    \multicolumn{13}{|c|}{Two-exponential model} \\ \hline
    Channel & $\chi_{\nu}^2$ & $p$ &$C_1$ & $\Delta C_1$ & $\tau_1$ [ns] & $\Delta \tau _1$ & $C_2$ & $\Delta C_2$ &$\tau_2$ [ns] & $\Delta \tau _2$ & $b$ & $\Delta b$ \\ \hline
    1 & 1.177 & < 10$^{-6}$ & 668 700 & 2 400 & 0.46018 & 0.00055 & 609 & 11 & 5.141 & 0.062 & 81.90 & 0.20 \\ \hline
    2 & 1.164 & < 10$^{-6}$ & 1 356 300 & 3 600 & 0.45550 & 0.00042 & 1 389 & 31 & 3.358 & 0.040 & 153.61 & 0.27 \\ \hline
    \multicolumn{13}{|c|}{Nonexponential model} \\ \hline
    Channel & $\chi_{\nu}^2$ & $p$ & $C$ & $\Delta C$ & $\tau$ [ns] & $\Delta \tau$ [ns] & $C_p$ & $\Delta C_p$ & $\beta$ & $\Delta \beta$ & $b$ & $\Delta b$ \\ \hline
    1 & 1.074 & 0.0035 & 693 300 & 2 600 & 0.44778 & 0.00062 & 2 988 & 76 & 1.559 & 0.014 & 75.54 & 0.30 \\ \hline
    2 & 1.042 & 0.0669 & 1 357 400 & 3 500 & 0.44717 & 0.00046 & 7 960 & 240 & 2.019 & 0.017 & 149.64 & 0.32 \\ \hline
    \end{tabular}
\end{table}

\begin{table}[!h]
    \centering
    \tiny
    \caption{Fitting results for Eosine Y measurements.}
    \label{data-analysis table 2}
    \renewcommand{\arraystretch}{1.7}
    \begin{tabular}{|c|c|c|c|c|c|c|c|c|c|c|c|c|}
    \hline
    \multicolumn{13}{|c|}{Fitting Range - Channel 1: 2.880 - 96.768 ns; Channel 2: 2.944 - 96.768 ns;} \\ \hline
    \multicolumn{13}{|c|}{Two-exponential model} \\ \hline
    Channel & $\chi_{\nu}^2$ & $p$ &$C_1$ & $\Delta C_1$ & $\tau_1$ [ns] & $\Delta \tau _1$ & $C_2$ & $\Delta C_2$ &$\tau_2$ [ns] & $\Delta \tau _2$ & $b$ & $\Delta b$ \\ \hline
    1 & 1.380 & < 10$^{-6}$ & 3 488 900 & 1 100 & 1.05587 & 0.00017 & 487.6 & 8.0 & 11.45 & 0.14 & 59.14 & 0.25 \\ \hline
    2 & 1.068 & 0.014 & 1 653 590 & 800 & 1.05545 & 0.00026 & 168 & 12 & 8.56 & 0.41 & 46.11 & 0.21 \\ \hline
    \multicolumn{13}{|c|}{Nonexponential model} \\ \hline
    Channel & $\chi_{\nu}^2$ & $p$ & $C$ & $\Delta C$ & $\tau$ [ns] & $\Delta \tau$ [ns] & $C_p$ & $\Delta C_p$ & $\beta$ & $\Delta \beta$ & $b$ & $\Delta b$ \\ \hline
    1 & 1.103 & 0.0005 & 3 483 200 & 1 200 & 1.05377 & 0.00017 & 7 860 & 360 & 1.512 & 0.019 & 48.13 & 0.53 \\ \hline
    2 & 1.028 & 0.1737 & 1 647 200 & 1 200 & 1.05480 & 0.00024 & 4 140 & 510 & 1.845 & 0.051 & 44.36 & 0.28 \\ \hline
    \end{tabular}
\end{table}

\section{Interpretation of the results}
\subsection{Shape of the spectral function close to threshold}
Turning to the interpretation of the data, we need to assess to what extent QM accounts for the observations\footnote{Delayed fluorescence occurs on much longer timescales and presumably cannot explain our data \cite{Croizat2020}.  Non-exponential fluorescence has also been reported in ensembles of distinct states as well as through singlet–triplet population recycling \cite{Wlodarczyk2004}; for a broader overview of these and other molecular processes distorting the decay rate, see Ref.~\cite{Medvedev1991}.}.
In molecular fluorescence, the relevant electronic states are $S_{0}$ (ground
singlet state) and $S_{1}$ (first excited singlet state) \cite{lakowicz}. 
After absorption
and rapid vibrational relaxation, the molecule resides in the lowest
vibrational level of $S_{1}$. This is a \textquotedblleft
narrow\textquotedblright\ initial state (Kasha's rule \cite{Kasha1950,IUPAC1997}). Fluorescence corresponds to the
radiative transition $S_{1}\;\rightarrow\ S_{0}$.

The existence of a single narrow state implies that a QM description in terms of the survival amplitude is possible:
\begin{equation}
    \label{A(t)}
    \mathcal{A}(t)=\int_{E_{th}}^{\infty} \!dE\, \rho(E)\, e^{-iEt/\hbar},
\end{equation}
where $\rho(E)$ is the energy distribution of the unstable quantum state and $E_{th}$ its minimal energy\footnote{In QFT one has similar features upon substituting $E \rightarrow s=E^2$, leading to   $\mathcal{A}(t)=\int_{s_{th}}^{\infty} \!ds\, \rho(s)\, e^{-i\sqrt{s}t/\hbar}$.}.  
The survival probability $P(t)=|\mathcal{A}(t)|^2$ satisfies $P(0)=1$, in turn implying $\int_{E_{th}}^{\infty} dE\, \rho(E)=1$. Then, the probability that a decay occurs between $t$ and
$t+dt$ is $-P'(t)dt$, implying that:
\begin{equation}
    I(t) \propto -P'(t) \text{ .}
\end{equation}

For a Breit-Wigner distribution  
\begin{equation}
\rho(E)=\frac{\Gamma}{2\pi}((E-M)^2+\Gamma^2/4)^{-1} \text{ ,}
\end{equation}
the exponential law(s) $P(t)=e^{-\Gamma t}$ and $I(t) \propto \Gamma e^{-\Gamma t}$ emerge. 
For realistic systems, $\rho (E)$ falls off faster than $E^{-2}$, thus $P'(0)=0$. 
The presence of a finite threshold $E_{th}$ is responsible for the long-time power law.

The late-time decay can be described by the following effective model \cite{ghirardi}:
\begin{equation}
\rho(E)=\mathcal{N}\,\frac{(E-E_{th})^{\gamma}}{(E-M)^{2}+\Gamma ^{2}/4}\,\theta(E-E_{th}) \text{ ,}
\label{sf}
\end{equation}
where $\mathcal{N}$ normalizes $\mathcal{A}(0)=1$, $\tau=\hbar/\Gamma$, and $M$ is the energy of the unstable state.
The exponent $\gamma$ determines the late-time scaling: 
$A(t)\sim t^{-(\gamma+1)}$, hence $P(t)\sim t^{-2(\gamma+1)}$ and $I(t)\sim t^{-(2\gamma+3)}=t^{-\beta}$ (thus, $\beta=2\gamma+3$).
The difference $M-E_{th}$ controls the turnover time: the smaller it is, the earlier the power-law sets in. 
Normalizability demands $|\gamma|<1$; 
for $-1<\gamma<0$ the spectral function diverges at threshold but remains integrable, leading to $1<\beta<3$. 
Our measured values of $\beta$ lie precisely in this range.
Thus, quantum effects can explain the measured power law, \textit{provided that} the density near threshold behaves as $\rho(E)\propto E^{\gamma<0}$. 

\begin{figure}[!h]
    \centering
    \begin{subfigure}{0.48\textwidth}
        \includegraphics[scale=1.2]{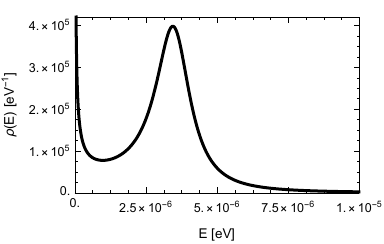}
    \end{subfigure}   
    \hfill
    \begin{subfigure}{0.48\textwidth}
        \includegraphics[scale=1.2]{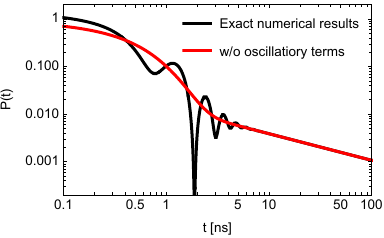}
    \end{subfigure}
    \caption{Spectral function $\rho(E)$ of Eq.(\ref{sf}) (left) and the corresponding survival probability (right), where oscillations appear (the superimposed curve is the coarse-grained outcome without oscillations ($t \gtrsim 2t_0$)).  The numerical values are chosen to reproduce the turnover and the late-time power law of channel 1 of Fig. \ref{fig:non-exp_comb}, erythrosine B case. }
    \label{fig:spectral-function}
\end{figure}

For the specific case of erythrosine B, channel 1, the appropriate spectral function and the survival probability are shown in Fig. \ref{fig:spectral-function}. 
Interestingly, $P(t)$ also shows pronounced oscillations due to the distorted threshold\footnote{Note, while the presence of oscillations is general, the specific amplitude depends on the employed model.}. 
In the physical case, these oscillations are washed out by ensemble averaging and the finite time-binning of detection, but they remain an intriguing QM prediction for future high-resolution studies. 

Interestingly, the values of $1<\beta<3$ (for which $\gamma<0$) are also listed in Ref. \cite{Rothe:2006rma}. (Other interpretations of these results, in terms of spectral filtering effects \cite{Martorell2007} or sample disorder \cite{Carminati2007}, have also been discussed.) Thus, a similar shape of the spectral function as in Fig. \ref{fig:spectral-function}, left panel, would also be required to describe the results in \cite{Rothe:2006rma}.

It remains an open question which physical mechanisms--whether related to the laser pulse generation or to internal molecular interactions--could be responsible for such a shape. If it originates from the state formation process, then altering the preparation conditions should modify the result, a possibility that is experimentally testable. If it arises from mutual interactions, a description via the statistical operator is required in future studies. In both cases, Eq. (\ref{sf}) shall be regarded as an effective description of the system.

\subsection{Multichannel decay at late times}
The two detectors measure the same exponential lifetime $\tau_1$ but different power law coefficients, see Table \ref{data-analysis table 3}. Now, for the QM interpretation to be attainable, it should, at least qualitatively, explain this result. 
To this end, we note that within the aforementioned transition $S_{1}\;\rightarrow\ S_{0}$,
the final state is not unique: the electron may decay into any of the many vibrational sublevels of
$S_{0}$ , with relative intensities determined by Franck--Condon factors \cite{lakowicz,Khanna2024}. Each such transition emits a photon with energy $E_{\gamma}%
=E(S_{1},v=0)-E(S_{0},v_{i}),$ where $i$ labels a level in $S_{0}%
$  \cite{Zhuang2011,Manian2021}. 

The relative intensity at each photon energy is given by the overlap of wave functions (the
emission being stronger for larger overlaps). The total decay width of the
excited state is the sum of the partial widths into all available channels,
$\Gamma_{\text{tot}}=\sum_{i}\Gamma_{i},$ with each $\Gamma_{i}$ corresponding
to decay into one specific level.
In the case of exponential decay, the intensity of a given decay channel is given by $I_i(t) \propto (\Gamma_i/\Gamma)e^{-\Gamma t}$, see e.g. \cite{Giacosa:2011xa,delaMadrid:2015ooa}.
Then, no band/channel difference takes place. This is the case of acridine orange. 

Yet, the picture changes when non-exponential deviations are accounted for. 
The theory for multichannel decay was derived by one of us (F. G.) \cite{Giacosa:2021hgl,Giacosa:2011xa}. In order to keep the discussion as simple as possible, let us consider only two channels. For the fluorescence system, they `roughly' represent the two detector bands (clearly, this is a
simplification since each band contains multiple channels, but the main
outcome can be generalized to more realistic cases). 
The total
spectral function of Eq. (\ref{sf}) is modified to include the sum of two channels, $\rho(E)=\rho_{1}(E)+\rho_{2}(E)$,
with
\begin{equation}
\rho_{i}(E)=\mathcal{N}\frac{c_{i}\left(  E-E_{th,i}\right)  ^{\gamma_{i}}%
}{(E-M)^{2}+\Gamma^{2}/4}\text{ ,}%
\end{equation}
where $E_{th,i}$ is the energy threshold of $i$-th channel. 
The probability that the decay takes place in the $i$-th channel between the time interval $(0,t>0$) reads \cite{Giacosa:2021hgl}:
\begin{equation}
w_{i}(t)=\int_{E_{th,i}}^{\infty}dE\frac{\Gamma_{i}(E)}{2\pi}\left\vert
\int_{0}^{t}\mathcal{A}(t^{\prime})e^{iEt^{\prime}/\hbar}dt^{\prime}\right\vert ^{2}\text{ ,
}\Gamma_{i}(E)=c_{i}\left(  E-E_{th,i}\right)  ^{\gamma_{i}}.
\end{equation}
For sufficiently large $t$, when the the power-law domain sets in and for a large time interval, $w_i(t)$ can be approximated by\footnote{Using $\int_{0}^{t}\mathcal{A}(t^{\prime})e^{iEt^{\prime}/\hbar}dt^{\prime} = \mathcal{G}(E)-\int_{t}^{\infty}\mathcal{A}(t^{\prime})e^{iEt^{\prime}/\hbar}dt^{\prime}$ where $\mathcal{G}(E)$ is the propagator, the scaling of Eq. (\ref{parti}) appears when the modulus squared is approximated as: $\left\vert ... \right\vert^2  \approx \left\vert\mathcal{G}(E) \right\vert^2-\mathcal{G}^*(E)\int_{t}^{\infty}\mathcal{A}(t^{\prime})e^{iEt^{\prime}/\hbar}dt^{\prime}-h.c.$.
}:
\begin{equation}
w_{i}(t)=w_{i}(\infty)-a_{i}t^{-(\gamma_{1}+\gamma_{2}+2)}-b_{i}%
t^{-2(\gamma_{i}+1)}-... \text{ ,}
\end{equation}
where $w_{i}(\infty)=\int_{E_{th,i}}^{\infty}dE\rho_{i}(E) \approx \Gamma_i/\Gamma$ is the
branching ratio for the $i$-channel, and $a_{i}$ and $b_{i}$ are appropriate coefficients\footnote{Interestingly, Eq. (\ref{parti}) can be also obtained by using the approximate expressions of Ref. \cite{Giacosa:2011xa}: $w_{i}%
(t)=w_{i}(\infty)-\operatorname{Re}[\mathcal{A}(t)\mathcal{A}^*_{i}(t)]$ with
$\mathcal{A}_{i}(t)=\int_{E_{th,i}}^{\infty}dE\rho_{i}(E)E^{-iEt/\hbar}.$ }.  
For definiteness, let us consider $\gamma_{1}<\gamma_{2}.$ Then, the distinct channel intensities at late times are:
\begin{equation}
I_{1}(t)\propto w_{1}^{\prime}(t)\propto t^{-2\gamma_{1}+3}\text{ ; }I_{2}(t) \propto
w_{2}^{\prime}(t)\propto t^{-(\gamma_{1}+\gamma_{2}+3)}\text{ .}%
\label{parti}
\end{equation}
This is a quite remarkable property: the late-time behaviour is at the onset of the non-exponential behaviour channel-dependent. This feature is
general and can be applied to any QM or QFT decay at late times, provided that at least two channels are present. The powers depend solely on the behaviour of the partial spectral
functions at thresholds (the parameters $\gamma_{i}$), hence the turnover times may
also differ. 

The channel dependence may be easily extended to bands, since each band is the sum over many channels. 
Hence, the different power laws for the
different detectors, see Fig. \ref{fig:non-exp_comb} and Table \ref{data-analysis table 3}, can be understood. As already stressed, this is not the case for a double exponential, for which the value of $\tau_2$ needs to be the same. 

The inclusion of different thresholds, the transition to a continuous model, and the onset of power tails are left as an outlook. They, however, do not change the main result that at late times the power law is channel (or band) dependent.

\section{Conclusions} 
In this work, we have shown that the fluorescence decay of erythrosine B and eosine Y are well described by power law curves (Fig. \ref{fig:non-exp_comb} and Tables \ref{data-analysis table 3}, \ref{data-analysis table 2}). For this to be a QM non-exponential decay as in Ref. \cite{ghirardi}, the spectral function must have a peculiar behaviour close to the left-threshold to reproduce the measured intensity (Fig. \ref{fig:spectral-function}). 
Moreover, based on Ref. \cite{Giacosa:2021hgl} and extending it, we have shown a general feature of multichannel quantum decays (of any type, thus valid in principle for any quantum unstable state, from elementary particles to molecular compounds):  contrary to the exponential decay with $\Gamma=\tau^{-1}$ common to any decay channel, the coefficient of the power law \textit{depends} on the specific channel (or, more generally, the band). Roughly speaking, since the power-law tail reflects a memory that the quantum system retains of its creation time, this memory is channel-dependent, and therefore `frequency' (or `colour') dependent. This property is consistent with the different power law coefficients that we measured. 

Future work on late-time multichannel decay is needed, both theoretically and experimentally, due to the complex nature of the system. Besides fluorescence, late-time deviations could also be investigated in the context of quantum tunnelling~\cite{dicus2002,Peshkin:2014jdw}, as e.g. the asymmetric potential well with two decay channels (left/right) of Ref.~\cite{Giacosa:2019jxz}. Applications to other systems, such as nuclear decays or even elementary particles, may also offer valuable opportunities to test these quantum predictions.

\bigskip

\textbf{Acknowledgments}
We thank M. Ga\'zdzicki, M. Pajek, P. Moskal, M. Skurzok, G. Pagliara, and S. Sharma for useful discussions. We also thank the Institute of Biology of the Jan Kochanowski University, in particular the head of the medical biology department, M. Arabski, for access to the facility.
This work was supported by the Minister of Science (Poland) under
the ‘Regional Excellence Initiative’ program (projects no. RID/SP/0015/2024/01 and no. RID/2025/LIDER/02).

\appendix
\section*{Appendix - Details of the fit}

\setcounter{figure}{0}
\renewcommand{\thefigure}{A\arabic{figure}}
\setcounter{table}{0}
\renewcommand{\thetable}{A\arabic{table}}
\setcounter{equation}{0}
\renewcommand{\theequation}{A.\arabic{equation}}

\FloatBarrier

\noindent 
In the experiment, we acquired data for various chemical compounds. The most promising for a~power law are eosine Y, and erythrosine B, see Tab. \ref{data-analysis table 3}, \ref{data-analysis table 2} in the main text. 
Acridine orange is reported in Tab. \ref{data-analysis table 4} as an example of a favoured two-exponential fit.  
In particular, we performed the following  measurements:
\begin{itemize}
    \item for erythrosine B: 9 measurements series (7 used in the analysis) obtained in three different trials (October 2024, November 2024, and May 2025);
    \item for eosine Y: 3 measurement series from December 2025;
    \item for acridine orange: 3 measurement series from October 2024.
\end{itemize}

\begin{figure}[!htbp]
    \centering

    \begin{subfigure}{0.45\textwidth}
        \includegraphics[width=\linewidth]{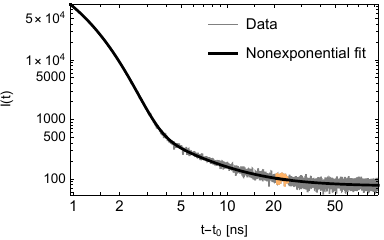}
        \caption{Erythrosine B, Channel 1}
        \label{fig:a}
    \end{subfigure}
    \hfill
    \begin{subfigure}{0.45\textwidth}
        \includegraphics[width=\linewidth]{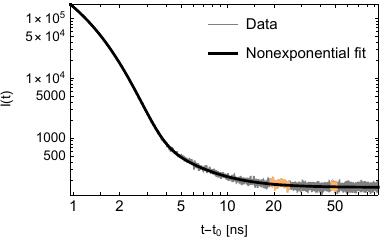}
        \caption{Erythrosine B, Channel 2}
        \label{fig:b}
    \end{subfigure}

    \vspace{2em}

    \begin{subfigure}{0.45\textwidth}
        \includegraphics[width=\linewidth]{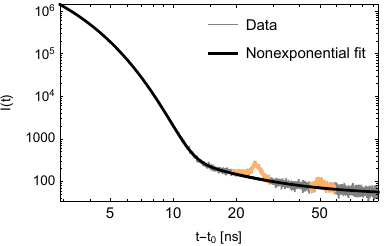}
        \caption{Eosine Y, Channel 1}
        \label{fig:c}
    \end{subfigure}
    \hfill
    \begin{subfigure}{0.45\textwidth}
        \includegraphics[width=\linewidth]{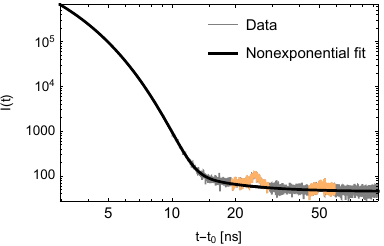}
        \caption{Eosine Y, Channel 2}
        \label{fig:d}
    \end{subfigure}

    \caption{Fluorescence intensity on log-log scale - full fitting range. Non-fitted data is marked with orange colour.}
    \label{fig:2x2panel-fluorescence-intensity-full-range}
\end{figure}

For each compound, the values from individual data sets were later combined into one table to achieve better statistics. The full datasets for erythrosine and eosine are reported in Fig. \ref{fig:2x2panel-fluorescence-intensity-full-range}.

The multi-exponential and QM-inspired power-law fitting functions, introduced in Tab. \ref{tab:my_label}, have been fitted to data using the standard $\chi ^2$ function defined as:
\begin{equation}
    \chi^2 = \sum _{n=1}^N \frac{(I_i-I(t_i))^2}{\sqrt{I_i}} \; ,
\end{equation}
where $N$ is the number of data points (in our experiment it is typically about 3000), $I_i$ denotes the measured number of counts in the $i$-th bin and $I(t_i)$ is the value of the number of counts in that bin predicted by the fitting function. Uncertainties corresponding to each bin are taken as the square root of the measured number of counts in that bin, appropriate for Poisson counting statistics. 
The results for the particular coefficients are presented in Tabs. \ref{data-analysis table 3}, \ref{data-analysis table 2}, and \ref{data-analysis table 4}. For the nonexponential QM-based fitting function, the values of the reduced $\chi_{\nu} ^2$, defined as $\chi_{\nu} ^2 = \chi^2 /df$, 
where $df$ denotes the number of degrees of freedom, are systematically lower than those corresponding to the standard multi-exponential model. In the following plots the full data sets are depicted; the agreement of the power-law extends over all times. 
\medskip

Standard methods based on the covariance matrix were employed. First, we construct the Hesse matrix $H$ whose elements are defined as second-order partial derivatives of the $\chi ^2$ function evaluated at the minimum:
\begin{equation}
    \label{Heesse_matrix}
    H_{ij} = \frac{\partial ^2 \chi^2}{\partial p_i\partial p_j} \; , 
\end{equation}

\noindent where $p_i$ denotes the $i$-th parameter of the fitting function ($i \in \{1,2,3,...,N \}$). In the next steps, the so-called covariance matrix $\Sigma$  is identified as the inverse of the Hessian: $\Sigma _{ij} = 1/2 \;\left( H^{-1} \right)_{ij}$.
The errors for the parameters follow as $\sigma_{p_i} = \sqrt{\Sigma_{ii}}$.
 The off-diagonal elements correspond to the covariances for all possible pairs formed from the set of the fit parameters. The example of such a covariance matrix for one particular case (nonexponential fit for the erythrosine B data, Channel 1) is explicitly presented below. The ordering of the parameters remains the same as in the last column of Tab. \ref{tab:my_label}.

\begin{equation*}
\Sigma=
    \left(
\begin{array}{ccccc}
3.27\times10^{6} & -0.735 & 3.04\times10^{4} & 4.44 & 42.4 \\
-0.735 & 1.95\times10^{-7} & -0.0142 & -2.30\times10^{-6} & -2.78\times10^{-5} \\
3.04\times10^{4} & -0.0142 & 2.92\times10^{3} & 0.520 & 7.40 \\
4.44 & -2.30\times10^{-6} & 0.520 & 9.85\times10^{-5} & 1.56\times10^{-3} \\
42.4 & -2.78\times10^{-5} & 7.40 & 1.56\times10^{-3} & 0.0451 \\
\end{array}
\right)
\end{equation*}

\begin{table}[!htbp]
        \centering
        \tiny
        \caption{Fitting results for Acridine Orange measurements.}
        \label{data-analysis table 4}
        \renewcommand{\arraystretch}{1.7}
        \begin{tabular}{|c|c|c|c|c|c|c|c|c|c|c|c|c|}
        \hline
        \multicolumn{13}{|c|}{Fitting Range - Channel 1: 3.200 - 96.768 ns; Channel 2: 3.200 - 96.768 ns;} \\ \hline
        \multicolumn{13}{|c|}{Two-exponential model} \\ \hline
        Channel & $\chi_{\nu}^2$ & $p$ &$C_1$ & $\Delta C_1$ & $\tau_1$ [ns] & $\Delta \tau _1$ & $C_2$ & $\Delta C_2$ &$\tau_2$ [ns] & $\Delta \tau _2$ & $b$ & $\Delta b$ \\ \hline
        1 & 1.098 & 0.0001 & 278 960 & 210 & 1.7335 & 0.0012 & 6355 & 60 & 5.953 & 0.019 & 21.99 & 0.12 \\ \hline
        2 & 1.042 & 0.0575 & 150 900 & 160 & 1.7326 & 0.0021 & 8984 & 64 & 5.949 & 0.016 & 40.58 & 0.16 \\ \hline
        \multicolumn{13}{|c|}{Nonexponential model} \\ \hline
        Channel & $\chi_{\nu}^2$ & $p$ & $C$ & $\Delta C$ & $\tau$ [ns] & $\Delta \tau$ [ns] & $C_p$ & $\Delta C_p$ & $\beta$ & $\Delta \beta$ & $b$ & $\Delta b$ \\ \hline
        1 & 11.088 & $ < 10^{-100}$ & 219 830 & 310 & 1.9473 & 0.0011 & 38 990 & 240 & 1.7853 & 0.0024 & -2.62 & 0.19 \\ \hline
        2 & 15.318 & $< 10^{-100}$ & 93 780 & 200 & 2.2810 & 0.0021 & 39 640 & 180 & 1.6773 & 0.0019 & 1.02 & 0.27 \\ \hline
        \end{tabular}
\end{table}

In connection to Fig. \ref{fig:2x2panel-fluorescence-intensity-full-range}, there are small but visible structures at 25 and 50 ns, with very small impact on the fit. These are an artifact due to the operation of the Time-to-Digital-Converter. The clock of this module of the TCSPC device operates with an internal reference clock of frequency 40 MHz, which corresponds exactly to 25 ns. Even a small nonlinearity in timing interpolation leads to slight modulations of the final signal at the observed times. This periodicity also represents strong evidence against other effects, such as afterpulsing.

Finally, we mention that also the IRF (Instrument Response Function) was measured. Subsequently, the convolution of the signal was performed, but the influence of the IRF turned out to be negligible for our late-time study.

\FloatBarrier

\bibliographystyle{unsrt}    
\bibliography{bibliography}

\end{document}